\documentclass[11pt]{article}
\usepackage{latexsym}
\usepackage{amssymb}
\usepackage[dvips]{graphicx}
\usepackage{epsfig}
\usepackage{rotating}
\usepackage{amsfonts}
\usepackage{amsmath}
\begin{document}

\begin{titlepage}
\begin{flushright}
NITheP-08-14\\
ICMPA-MPA/2009/10\\
\end{flushright}

\begin{center}

{\Large\bf Coherent states in noncommutative quantum mechanics}

J Ben Geloun$^{a,b,c,*}$ and F G Scholtz$^{a,\dag}$

$^{a}${\em National Institute for Theoretical Physics}\\
{\em Private Bag X1, Matieland 7602, South Africa}\\
$^{b}${\em International Chair of Mathematical Physics
and Applications}\\
{\em ICMPA--UNESCO Chair 072 B.P. 50  Cotonou, Republic of Benin}\\
$^{c}${\em D\'epartement de Math\'ematiques et Informatique}\\
{\em  Facult\'e des Sciences et Techniques, Universit\'e Cheikh Anta Diop, Senegal}

E-mail:  $^{*}$bengeloun@sun.ac.za,\quad $^{\dag}$fgs@sun.ac.za  

\today

\begin{abstract}
\noindent 
Gazeau-Klauder coherent states in noncommutative
quantum mechanics are considered. We find that these states share
similar properties to those of ordinary canonical coherent states in the sense that they 
saturate the related position uncertainty relation, obey 
a Poisson distribution and possess a flat geometry. 
Using the natural isometry between the quantum 
Hilbert space of Hilbert Schmidt operators and the tensor product
of the classical configuration space and its dual,
we reveal the inherent vector feature of these states.
\end{abstract}

\end{center}

Pacs number 11.10.Nx, 03.65.-w

\end{titlepage}

\setcounter{footnote}{0}

In the search of a unifying theory of gravity and quantum 
mechanics, noncommutative geometry is beginning to play an increasing role \cite{dop,doug}.
Recently, a formulation of noncommutative quantum 
mechanics, which allows for a consistent interpretation of position measurement  \cite{sch}
and the solution of the difficult problem of a noncommutative well \cite{sj}
has been put forward. A key ingredient in this construction is a coherent
state on the quantum Hilbert space (space of Hilbert Schmidt operators on the classical 
configuration space \cite{sch}), which is expressed in terms of a projection
operator on the usual Glauber-Klauder-Sudarshan coherent states in the classical configuration space. Using these coherent states a positive operator valued measure 
was constructed and used to give a probabilistic interpretation to position measurement \cite{sch}. In \cite{sch2}, these states were used to derive a path integral
representation for the propagator of the free particle. The result of this
procedure is the appearance of an UV cut-off, which settled a long standing dispute \cite{sch2}. These results motivate for a systematic study of this class of 
coherent states in quantum Hilbert space. 
Let us mention that in previous analyses \cite{may}-\cite{yin} coherent states
over noncommutative spaces have been discussed through deformed
Heisenberg-Weyl algebras. In \cite{yiu}, for instance, the coherent states
have been built from non diagonal bosonic operators. 

However, we do not proceed in the same way, following rather the approach of 
\cite{sch} and furthermore investigate how to implement these coherent states 
with the physical axioms of Gazeau-Klauder \cite{gk,kl2}. 
Hence, in this letter, we built Gazeau-Klauder coherent
states on the quantum Hilbert space spanned by the eigenvectors
of a Hamiltonian operator. 
Displacement operators and mathematical 
properties are also discussed. Finally, mapping isometrically the quantum Hilbert space onto a tensor product of the configuration space and its dual,
we interpret these states as vector coherent states in the sense of Ali {\it et al} \cite{al}.

To begin with, let us consider the two dimensional noncommutative coordinate algebra 
\begin{equation}
[\hat{x}, \hat{y} ]=i\,\theta,
\label{ncb}
\end{equation}
with the parameter $\theta$ referred to as the noncommutative parameter.
The annihilation and creation operators 
$b=(1/\sqrt{2\theta})(\hat{x} + i \hat{y})$ 
and $b^\dagger=(1/\sqrt{2\theta})(\hat{x} - i \hat{y})$ obey
a Heisenberg-Fock algebra $[b,b^\dag]=\,1\!\!1_c$.
Hence the noncommutative configuration space ${\cal H}_c$ becomes itself a Hilbert space 
isomorphic to the boson Fock space ${\cal H}_c = {\rm span}\{|n\rangle, n\in \mathbb{N}\}$, with $|n\rangle = (1/\sqrt{n!})(b^\dag)^{n}|0\rangle$.  
 
At the quantum level, the Hilbert space ${\cal H}_q$ representing the states of the system, is defined to be the space of Hilbert-Schmidt operators acting on ${\cal H}_c$ \cite{hol}: 
\begin{equation}
\label{qhil}      
\mathcal{H}_q = \left\{ \psi(\hat{x}_1,\hat{x}_2): \psi(\hat{x}_1,\hat{x}_2)\in \mathcal{B}\left(\mathcal{H}_c\right),\;{\rm tr}_c(\psi(\hat{x}_1,\hat{x}_2)^\dagger\psi (\hat{x}_1,\hat{x}_2)) < \infty \right\},
\end{equation}
where ${\rm tr}_c$ denotes the trace over ${\cal H}_c$ and
$\mathcal{B}\left(\mathcal{H}_c\right)$ is the set of bounded operators
on $\mathcal{H}_c$.
Next, we introduce the noncommutative Heisenberg algebra\footnote{
Henceforth capital letters refer to quantum operators acting on ${\cal H}_q$ in order to distinguish them from classical operators acting on ${\cal H}_c$.} 
\begin{eqnarray} 
&&\begin{array}{rcl}
&&[\hat{X},  \hat{Y} ] = i\theta, \\
&&[\hat{X}, \hat{P}_{X} ] = i\hbar=
[\hat{Y}, \hat{P}_{Y} ],  \\
&&[\hat{P}_{X}, \hat{P}_{Y} ]= 0,
\end{array}
\label{qbalg}
\end{eqnarray} 
of which we seek a representation on ${\cal H}_q$.
It can be verified that the following relations provide a well defined representation 
\begin{eqnarray} 
&&\hat{X} \psi= \hat{x} \psi, \quad
\hat{Y} \psi =  \hat{y}\psi , \quad 
\hat{P}_{X} \psi =\frac{\hbar}{\theta} [\hat{y},\psi], \quad
\hat{P}_{Y} \psi= -\frac{\hbar}{\theta} [\hat{x},\,\psi],
\label{brep}
\end{eqnarray}
with self-adjoint properties with respect to the quantum Hilbert space inner product $(\phi|\psi)={\rm tr}_c (\phi^\dag \psi)$. Thus, (\ref{brep}) determines a unitary representation.

Coherent states in the classical Hilbert space ${\mathcal H}_c$
are well known to be
\begin{eqnarray}
|z\rangle = e^{-\frac{|z|^2}{2}} \sum_{n=0}^{\infty}
\frac{z}{\sqrt{n !}}\;  |n\rangle.
\end{eqnarray}
Now we equip such states with a parameter $\tau$ such that they
fulfill the Gazeau-Klauder axiom of temporal stability relative to the 
classical time evolution operator
\begin{eqnarray} 
&&U(t) = e^{-i t \omega_0 H^{\rm red}},\qquad
H^{\rm red}=\frac{H}{E_0} = \frac{1 }{2\theta} (\hat x^2  + \hat y^2),\\
&&H^{\rm red}= e_n |n\rangle, \quad e_n = n + \frac{1}{2},
\end{eqnarray}
where $E_0=\hbar\omega_0$ is some energy scale.
The set of coherent states
\begin{eqnarray}
|z, \tau \rangle = e^{-\frac{|z|^2}{2}} \sum_{n=0}^{\infty}
\frac{z^n}{\sqrt n !} e^{-i \omega_0 \tau e_n}|n\rangle,
\label{cs1}
\end{eqnarray}
transform into one another under time translation, namely,
\begin{eqnarray}
U(t) |z, \tau \rangle =  |z, \tau +t \rangle.
\end{eqnarray} 
The set of states (\ref{cs1}) are continuous in the label $z$, normalized to unity,
stable under time evolution and finally form an overcomplete set
in the Hilbert space ${\mathcal H}_c$, i.e. they solve
\begin{eqnarray}
\,1\!\!1_c= \sum_{n=0}^{\infty}\,|n\rangle\, \langle n|\,=\, \int_{\mathbb{C}} \frac{dz}{\pi} |z, \tau \rangle \langle z, \tau|,
\end{eqnarray} 
and subsequently, fulfill the Gazeau-Klauder axioms for coherent states.

On the quantum level, we also need to fix an evolution
operator for a given Hermitian Hamiltonian operator $H_q$ on the quantum Hilbert
space ${\mathcal H}_q$. 
To proceed, we consider a quantum system with a diagonalizable Hamiltonian operator 
$H_q$ admitting the spectral decomposition 
 \begin{eqnarray}
 H_q  = \sum_{n,m=0}^{\infty} |\Psi_{n,m})\, \widetilde{E}_{n,m}\, (\Psi_{n,m}|,
 \end{eqnarray}
with respect to an orthonormal eigenstate basis $\{|\Psi_{n,m}), \;n,\,m \in \mathbb{N} \}$, i.e. $(\Psi_{n,m}|\Psi_{n',m'})= {\rm tr}_c (\Psi_{m,n}^\dag\Psi_{n',m'})$ $=\delta_{m,m'}\delta_{n,n'}$. 
We assume that the basis $\{|\Psi_{n,m})\}$ is related 
to the orthonormal quantum Hilbert space basis $\{|n\rangle\langle m|=|n, m)\}$ via a unitary transformation: 
\begin{eqnarray} 
{\mathcal U} |n, m) = |\Psi_{n,m}),\quad
{\mathcal U}^\dag |\Psi_{n,m})= |n, m).
\end{eqnarray}
An expansion of these operators is given by
\begin{eqnarray} 
{\mathcal U} = \sum_{n,m}^{\infty}\; 
|\Psi_{n,m}) \,(n, m|, 
\qquad
{\mathcal U}^\dag =\sum_{n,m}^{\infty}\; |n,m)\, (\Psi_{n,m}|.
\end{eqnarray}
One verifies that
\begin{eqnarray} 
{\mathcal U} {\mathcal U}^\dag = 
\sum_{n,m}^{\infty}\; |\Psi_{n,m}) \, (\Psi_{n,m}| = \mathbb{I}_q,\qquad
  {\mathcal U}^\dag {\mathcal U} =
  \sum_{n,m}^{\infty}\; |n,m)\,\, (n,m|
  = \mathbb{I}_q,
  \end{eqnarray}
where $\mathbb{I}_q$ stands for the quantum Hilbert space identity.
For instance, if  $H_q$ is the noncommutative quantum Harmonic oscillator
Hamiltonian, which proves to be solvable \cite{sch}, the unitary operators
${\mathcal U}$ and ${\mathcal U}^\dag$ could be exactly specified. 
Nevertheless, let us keep this general formulation
for any diagonalizable noncommutative quantum system. 

An evolution operator on the space spanned by the basis 
$\{ |n, m)\}$ is therefore given by
\begin{eqnarray} 
&&\mathbb{U}(t) =e^{-it\omega_0 \mathbb{H}^{\rm red}},\\
&&\mathbb{H}^{\rm red}= {\mathcal U}^\dag \; H^{\rm red}_q\; {\mathcal U} = \sum_{n,m=0}^{\infty}\, |n, m)\,  E_{n,m}\,
 (n,m|,\;\, H^{\rm red}_q = H_q/(\hbar\omega_0),
\end{eqnarray}
where $H^{\rm red}_q$ is the dimensionless quantum Hamiltonian corresponding
to $H_q$ with associated spectrum $E_{n,m}= \widetilde{E}_{n,m}/(\hbar\omega_0)$. 
  
Consider the set of states defined as 
\begin{eqnarray}
|z,\tau) & =&  \mathbb{U}(\tau)|z \rangle \, \langle  z|  \cr
&=& e^{-|z|^2} \sum_{n,m} 
\frac{z^n \, \bar z^{m}}{\sqrt{n!\, m!}} e^{-i \omega_0 \tau E_{n,m}} |n,m).
\label{cs2}
\end{eqnarray}
Let us prove that the set of states (\ref{cs2}) are Gazeau-Klauder
coherent states. 

\noindent $\bullet$ The {\it normalizability} condition is verified since
\begin{eqnarray}
(z,\tau|z,\tau) = {\rm tr}_c 
((\mathbb{U}(\tau)|z \rangle \, \langle  z| )^\dag \mathbb{U}(\tau)|z \rangle \, \langle  z|)= (\langle  z|z \rangle)^2 = 1.
\end{eqnarray} 
$\bullet$ The {\it continuity in labeling} consists of the statement
\begin{eqnarray}
\forall z,z'\in \mathbb{C},\;\;\; |z-z'| \to 0,\qquad
|||z,\tau) -|z',\tau)  ||^2_{HS} \to 0,
\end{eqnarray}  
where the norm is that of Hilbert-Schmidt. Evaluating the second
member of the proposition, one has
\begin{eqnarray}
|||z,\tau) -|z',\tau)  ||^2_{HS} &=&
2- {\rm tr}_c (|z \rangle \, \langle  z|z' \rangle\langle  z'| +
|z' \rangle \langle  z'|z \rangle\langle  z|)\\
&=& 2(1- e^{-|z-z'|^2})
\end{eqnarray} 
which tends to zero whenever $|z-z'|$ is sufficiently small.
Thus the states (\ref{cs2}) are continuous in $z$.

\noindent $\bullet$ The {\it temporal stability} axiom can be
readily inferred. We have 
\begin{eqnarray}
\mathbb{U}(t) |z,\tau) = |z,\tau+ t).
\end{eqnarray}  

\noindent $\bullet$ The {\it overcompleteness relation} is a fundamental
property that each set of coherent states ought to satisfy. 
A proof of this condition is given in \cite{sch} and in \cite{sch2} where
the associated momentum coherent basis with a $\star$ product insertion was used.
In the subsequent developments, we formulate this proof
in the context of vector coherent states.
We claim the following statement 
\begin{eqnarray}
\int_{\mathbb C} \frac{dz d\bar z}{\pi} |z,\tau)\star (z,\tau| 
= \int_{\mathbb C} \frac{dz d\bar z}{\pi} |z,\tau)e^{\stackrel{\leftarrow}{\partial}_{\bar z} \stackrel{\rightarrow}{\partial}_z	} (z,\tau|= \mathbb{I}_q.
\label{resolu}
\end{eqnarray} 
Note that the set of quantum coherent states could be realized 
in terms of classical Gazeau-Klauder coherent states (\ref{cs1}). 
However, the temporal stability axiom could be settled only if
one removes the previous phase factor and then inserts 
a new time parameter, 
$$\widetilde{|z,\tau)} =\mathbb{U}(\tau) 
U(-\tau')|z,\tau'\rangle \langle z, \tau'|U(+\tau').$$   
Finally, in order to obtain a set of coherent states over
the initial Hamiltonian quantum Hilbert space generated by $H^{\rm red}_q$, 
one has to  map unitarily the coherent states (\ref{cs2}) via 
${\mathcal U}$. All the above statements then remain true under
this unitary transformation.
 
Let us investigate some basic properties of the coherent states
(\ref{cs2}). At first, we may ask if there is a unitary displacement
operator which could generate them. Consider the operators
\begin{eqnarray}
D_R = e^{z b^\dag - \bar z b}_{R},\quad
D_L = e^{-z b+ z b^\dag}_{L},\quad
\end{eqnarray} 
where the lower indices $R,L$ of the exponential operators
refer to right and left action, respectively. The composition
\begin{eqnarray}
D(\tau) =  \mathbb{U}(\tau) D_R D_L 
\end{eqnarray} 
acts on the vacuum as follows
\begin{eqnarray}
D(\tau)|0 \rangle \langle 0|&=&\mathbb{U}(\tau) e^{-|z|^2}
e^{z b^\dag }|0 \rangle \, \langle 0|e^{-z b }= |z,\tau),
\end{eqnarray} 
and verifies $D^\dag(\tau) D(\tau) = \mathbb{I}_q$. Therefore $D$
is a unitary displacement operator of coherent states (\ref{cs2}).

Statistical properties of these states can also be investigated.
Let us start by squeezing properties. The momentum operators read 
\begin{eqnarray}
\hat P_X \cdot = \frac{-i\hbar}{\sqrt{2\theta}}[(b-b^\dag)
\,,\, \cdot],\qquad
\hat P_Y \cdot= \frac{\hbar}{\sqrt{2\theta}}[(b+b^\dag)\,,\, \cdot] .
\end{eqnarray}
Variances of the operators $\hat X, \hat Y, \hat P_X$ and $\hat P_Y$
in any coherent state $|z,\tau)$ can be derived directly using the  
Barut-Girardello equation $b|z,\tau)= z|z,\tau)$. We obtain
\begin{eqnarray}
&&[\Delta \hat X]^2 = (\,\hat X^2\,) - (\,\hat X\,)^2 = \frac{\theta}{2},\\
&&[\Delta \hat Y]^2 = (\,\hat Y^2\,) - (\,\hat Y\,)^2 = \frac{\theta}{2},\\
&&[\Delta \hat P_X]^2 = (\,\hat P_X^2\,) - (\,\hat P_X\,)^2 = \frac{\hbar^2}{\theta},\\
&&[\Delta \hat P_Y]^2 = (\,\hat P_Y^2\,) - (\,\hat P_Y\,)^2 = \frac{\hbar^2}{\theta},
\end{eqnarray}  
giving the uncertainties
\begin{eqnarray}
&&[\Delta \hat X \,\Delta \hat Y]^2 =\frac{\theta^2}{4} = \frac{1}{4}|([\hat X, \,\hat Y])|^2,\\
&&[\Delta \hat X\Delta \hat P_X]^2 = \frac{\hbar^2}{2} \geq \frac{1}{4}|([ \hat X, \hat P_X])|^2,\\
&&[\Delta \hat Y\Delta \hat P_Y]^2 = \frac{\hbar^2}{2} \geq \frac{1}{4}|([ \hat Y, \hat P_Y])|^2,\\
&&[\Delta \hat P_X\Delta \hat P_Y]^2 = \frac{\hbar^4}{\theta^2} \geq  \frac{1}{4}
|([ \hat P_X, \hat P_Y])|^2=0.
\end{eqnarray}  
As expected, the first equality shows  that the coherent states are {\it intelligent} \cite{nieto}. In fact, the saturation of uncertainty relations relative to 
$(\hat X,\hat P_X)$ and $(\hat Y,\hat P_Y)$ could be derived only through 
coherent states in the momentum sector 
(an explicit expression of these state are given in \cite{sch2}). 
This is  also in contrast with previous results \cite{yiu} where coherent states
(built in noncommutative system) saturate only the uncertainty relations 
for (position,momentum) sectors. 

We pursue the statistical analysis and find the photon-number distribution of the
coherent states (\ref{cs2}) by setting the time parameter $\tau=0$, for simplicity. 
This is quite straightforward once 
the mean value of the monomial $(b^\dag)^p b^q$, $p,q \in \mathbb{N}$,
has been found. Indeed,
\begin{eqnarray}
(\,(b^\dag)^p b^q\,)_0 = (z,0| (b^\dag)^p b^q |z,0) = 
\langle z|\,(b^\dag)^p b^q\,|z\rangle = \bar z^p \, z^q.
\label{mon} 
\end{eqnarray} 
This proves according to the ordinary coherent states formulation
in the commutative limit, that they obey to the same photon-number 
Poisson distribution $P(x,n)= x^n e^{-x}/ n!$, with no deviation 
corresponding to a Mandel parameter \cite{man} $Q=1$. 
We can also infer from the above equation (\ref{mon}) that the geometry
of the coherent states (\ref{cs2}) coincides exactly with 
the flat geometry of ordinary commutative ones, with Fubini-Study element 
$d\sigma = W(|z|)dz d \bar z$ with a constant metric
factor $W(|z|)= 1$ \cite{field}.
 
In this last paragraph we sketch how the coherent states (\ref{cs2})
could be implemented with a vector character \cite{al}. 
We have the isomorphic isometry \cite{bar}
\begin{eqnarray}
{\mathcal H}_q \equiv {\mathcal H}_c \otimes {\mathcal H}_c^*,
\end{eqnarray}   
where ${\mathcal H}_c^*$ is the  dual Hilbert (isomorphic) space of the classical
space ${\mathcal H}_c$. Therefore a basis of ${\mathcal H}_q$ is 
of course given by $\{| n\rangle \otimes \langle m|,\, n,m \in \mathbb{N}\}$.
As a consequence, there is a natural inner product in ${\mathcal H}_q $
induced by the product of inner products of each its component
subspaces ${\mathcal H}_c$ and ${\mathcal H}_c^*$. Adjoint properties
could be also defined via this inherited inner product.

The set of bounded operators $F_n$, $\forall n \in \mathbb{N}$,
\begin{eqnarray}
&& F_n: \mathbb{C} \to {\mathcal B}({\mathcal H}_c)\\
&& F_n: z \;\mapsto\;\; F_n(z): {\mathcal H}_c \to {\mathcal H}_c\\
&& F_n(z) |m\rangle = \frac{z^n \bar z^m}{\sqrt{n!m!}} e^{-i\tau \omega_0 E_{n,m}} |m\rangle.
\end{eqnarray}
such that any $F_n(z)$ is continuous in $z$, allows to construct the set of vectors
\begin{eqnarray}
&&|z,\tau, m ) = N(|z|)^{-\frac{1}{2}} \sum_{n\in \mathbb{N}}   F_n(z) |m\rangle \otimes \langle n|
\label{vcs}\\
&& N(|z|) = \sum_{n=0}^{\infty} {\rm tr}_c |F_n(z)|^2 =e^{2|z|^2}< \infty.
\label{norm}
\end{eqnarray} 
The vectors (\ref{vcs}) could be also generated by the operator
$T:{\mathcal H}_c \to {\mathcal H}_c\otimes {\mathcal H}_c^*$, such that
$T|m\rangle = |z,\tau, m )$. The fact that  $T$ is bounded (as one can deduce
from (\ref{norm})) becomes central 
when one has to interchange infinite sums and integrals in the computations.

The set of vectors $\{|z,\tau, m ), m\in \mathbb{N}\}$ verifies the 
axioms of Ali {\it et al}:

\noindent $\bullet$ Normalization condition:
\begin{eqnarray} 
\sum_{m=0}^\infty  (z,\tau, m |z,\tau, m ) = 
N(|z|)^{-1}\sum_{n,m=0}^\infty \frac{|z|^{2m}|z|^{2n}}{n!m!} = 1.
\end{eqnarray} 
$\bullet$ Resolution of the identity
\begin{eqnarray} 
\mathbb{I}_q = 1\!\!1_c \otimes  1\!\!1_c^* =
\sum_{m=0}^\infty \frac{1}{m!}\int_{D} d\nu(z)\,\, (\stackrel{\rightarrow}{\partial}_{\bar z})^m\, \left\{\,\,
N(|z|)\,|z,\tau, m )\, 
( z,\tau, m | \,\,\right\}(\stackrel{\leftarrow}{\partial}_{z})^m,
\label{vcsres}
\end{eqnarray} 
where $d\nu(z)$ is a measure over some disc $D$ of $\mathbb{C}$ to be specified. 
Note also the particular place of the normalization factor $N(|z|)$.
The latter equation is of course an adapted version of (\ref{resolu}) for 
the vector states (\ref{vcs}) and extends Ali's formulation of the resolution 
of the identity for vector coherent states in noncommutative spaces. 

In order to prove (\ref{vcsres}), we first expand the integrand as
\begin{eqnarray}
&&(\stackrel{\rightarrow}{\partial}_{\bar z})^m\, \left\{\,\,
N(|z|)\,|z,\tau, m )\, 
( z,\tau, m | \,\,\right\}(\stackrel{\leftarrow}{\partial}_{z})^m\cr
&&=
\sum_{n,n'\in \mathbb{N}}  (\stackrel{\rightarrow}{\partial}_{\bar z})^m(F_n(z) |m\rangle \otimes \langle n|) 
( F_{n'}(z) |m\rangle \otimes \langle n'|)^\dag (\stackrel{\leftarrow}{\partial}_{z})^m \cr
&&:= 
\sum_{n,n'\in \mathbb{N}}  [(\partial_{\bar z})^m\,F_n(z) |m\rangle \langle m|(\partial_{ z})^m(F_{n'}(z))^*] \otimes |n \rangle \langle n'|.
\label{prim}
\end{eqnarray}  
Let us call $A$ the operator of the right hand side of (\ref{vcsres})
and then choose two vectors of ${\mathcal H}_q$, say 
$|\Psi)=\sum_{p,q=0}^\infty\Psi_{pq} |p \rangle \otimes \langle q|$
and $|\Phi)=\sum_{k,l=0}^\infty\Phi_{kl} |k \rangle \otimes \langle l|$,
with $\Psi_{pq}$ and $\Phi_{pq}$ complex numbers. 
By definition of weak convergence and using (\ref{prim}), we get
\begin{eqnarray} 
&&(\Psi| A |\Phi) =
\cr
&& 
\sum_{m=0}^\infty \frac{1}{m!}\int_{D} d\nu(z)
\sum_{n,n'\in \mathbb{N}}\,\sum_{k,l,p,q=0}\,
\Psi^*_{pq} \Phi_{kl} [(\partial_{\bar z})^m\,F_n(z) |m\rangle \langle m|(\partial_z)^n(F_{n'}(z))^*] \otimes |n \rangle \langle n'|\cr&&
\end{eqnarray} 
where the boundness of $T$ has been used to interchange the
external sums (over $p,q$ and $k,l$) and the integral with the sum over $m$.
Using the radial parametrization of the measure $ d\nu(z)= dz W(z)$, $z=re^{i \theta}$, $r\in [0,\infty)$, $\theta\in[0,2\pi[$,  we get
\begin{eqnarray}
(\Psi| A |\Phi)= \sum_{m=0}^\infty \sum_{n=0}^\infty \Psi^*_{mn} \Phi_{mn} \left\{
2\pi \int_{0}^\infty r dr W(r) \frac{r^{2n}}{n!}\right\} |m\rangle \langle m|
\otimes |n \rangle \langle n|,
\end{eqnarray}
where, again in view of the boundness of $T$, the
sums over $m$ and $n$ and the integral commute. The resulting momentum problem 
can be easily solved by $ W(r) = (1/\pi) e^{-r^2}$. One ends with 
the scalar product $(\Psi| A |\Phi) = (\Psi|\Phi)$ which achieves the proof 
of the resolution of the identity.

The vector coherent states (\ref{vcs}) verifies also the Gazeau-Klauder
axioms. Indeed, once a quantum Hamiltonian onto the tensor product
Hilbert space is given, all the Gazeau-Klauder axioms can be checked 
according to our previous analysis.
 
Finally, we emphasize that it is the measure used
to integrate the (vector) coherent states to unity, which is responsible
for the appearance of the $\star$-product. In fact, if one uses a measure 
$d\nu(z)$ as a tensor product of measures $d\nu(z)_1\otimes d\nu(z)_2$  over the spaces ${\mathcal H}_c$ and ${\mathcal H}^*_c$, it is straightforward to conclude that 
the states $\mathbb{U}(\tau)| z\rangle \otimes \langle z| $  
satisfy the Gazeau-Klauder properties. The resolution of the identity is simply
a product of resolutions of the identity in each space.
Therefore, it also seems possible to define (vector) coherent states avoiding
any occurrence of $\star$-product. This could be seen by the prescription
of a particular order 
$$|z,\tau)= \mathbb{U}(\tau)| z\rangle \wedge \langle z| =\mathbb{U}(\tau)
f\left(| z\rangle \otimes \langle z|\right),
$$
for which a resolution of the identity could reduce to the sum
of two resolution of the identities for each set of states such that
the ordinary measure is of the form 
$W(z,\bar z) dz \wedge d \bar z=W(z,\bar z)(dz\otimes d\bar z - d\bar z \otimes dz)
$. An external multiplication law between 
${\mathcal H}_c\otimes {\mathcal H}^*_c$ and 
${\mathcal H}^*_c\otimes {\mathcal H}_c$ has also to be specified 
in this context.

\section*{Acknowledgments}
This work was supported under a grant of the  National Research Foundation of South Africa.

\end{document}